\documentclass[aps,prd,superscriptaddress,nofootinbib,notitlepage]{revtex4-1}
\usepackage{amsmath,amssymb,amsfonts}
\usepackage{graphicx}
\usepackage{hyperref}
\usepackage{xcolor}
\usepackage{bbold}
\usepackage{bm}
\usepackage{algorithm}
\usepackage{algpseudocode}
\usepackage{subcaption}
\newcommand{\Tr}{\mathrm{Tr}}
\newcommand{\ReTr}{\mathrm{Re}\,\mathrm{Tr}}

\newcommand{\expect}[1]{\langle #1 \rangle}

\begin{document}

\title{Diffusion Models for SU(2) Lattice Gauge Theory in Two Dimensions}

\author{H.~Alharazin}
\affiliation{Institut f\"ur Theoretische Physik II, Ruhr-Universit\"at Bochum, D-44780 Bochum, Germany}

\author{J.~Yu.~Panteleeva}
\affiliation{Institut f\"ur Theoretische Physik II, Ruhr-Universit\"at Bochum, D-44780 Bochum, Germany}

\author{B.-D.~Sun}
\affiliation{Institut f\"ur Theoretische Physik II, Ruhr-Universit\"at Bochum, D-44780 Bochum, Germany}

\date{\today}

\begin{abstract}
We apply score-based diffusion models to two-dimensional SU(2) lattice pure gauge theory with the Wilson action, extending recent work on U(1) gauge theories.~The SU(2) manifold structure is handled through a quaternion parameterization.~The model is trained on 10,000 configurations generated via Hybrid Monte Carlo at a fixed coupling $\beta_0= 2.0$ on an $8\times 8$ lattice, augmented to 20,000 samples via random gauge transformations.~Through physics-conditioned sampling exploiting the linear $\beta$-dependence of the score function, we generate configurations at different values of the coupling without retraining;~through the fully convolutional U-Net architecture with periodic boundary conditions, we generate configurations on lattices of different spatial extents.~We validate our approach by comparing the average plaquette against exact analytical predictions.~At the training lattice size ($8\times 8$), the model reproduces the exact plaquette with biases $|\Delta| \leq 0.001$ for $\beta \in [1.5, 2.5]$ and $|\Delta| < 0.06$ across $\beta \in [1, 4]$.~For lattices sharing the training extent $L=8$ in at least one direction, biases remain below $\sim 0.003$ for $\beta \in [1.5, 2.5]$, with larger deviations at higher couplings.~This work demonstrates that diffusion models are a promising tool for non-Abelian gauge field generation and motivates further investigation toward higher-dimensional theories.
\end{abstract}

\maketitle

%=====================================================================
\section{Introduction}
\label{sec:introduction}
%=====================================================================

Lattice gauge theory provides a rigorous non-perturbative framework for studying quantum chromodynamics (QCD) and other gauge theories~\cite{Wilson1974,Creutz1980}.~The fundamental computational challenge lies in sampling gauge field configurations from the Boltzmann distribution $p(U) \propto e^{-S[U]}$, where $S$ denotes the Euclidean action.~Traditional Markov Chain Monte Carlo (MCMC) methods, particularly the Hybrid Monte Carlo (HMC) algorithm~\cite{Duane1987}, have been the workhorse of lattice QCD for decades.~However, these methods face severe limitations:~critical slowing down near continuum, topological freezing at fine lattice spacings \cite{DelDebbio:2002xa}, and the notorious sign problem for finite baryon density~\cite{deForcrand2010}.

The emergence of machine learning approaches offers promising avenues to address these challenges.~Normalizing flows have been successfully applied to scalar field theories~\cite{Albergo2019} and gauge theories~\cite{Kanwar2020,Boyda2021,Abbott2023}, demonstrating the potential of generative models in lattice field theory.~More recently, diffusion models---a class of generative models that learn to reverse a gradual noising process---have achieved remarkable success in image generation~\cite{Ho2020,Song2021} and have been proposed for lattice gauge theory applications~\cite{Wang:2023exq,Wang:2023sry,Zhu2024a,Zhu2024b,Vega:2025hgz,Kanwar:2025wuc,Aarts:2026zzr}.

The work of Zhu \textit{et al.}~\cite{Zhu2024a,Zhu2024b} established a diffusion model framework for two-dimensional U(1) gauge theory, demonstrating several attractive properties: the ability to generate configurations at different couplings through physics-conditioned sampling, generalization to different lattice sizes via a fully convolutional architecture, and avoidance of topological freezing.~A framework with these capabilities, training in a regime where HMC is efficient and extending to regions where it suffers from critical slowing down or topological freezing, is particularly promising and warrants further investigation.~However, the extension to non-Abelian gauge theories, essential for QCD applications, remained unexplored.~During the preparation of this manuscript, Aarts \textit{et al.}~\cite{Aarts:2026zzr} addressed this extension using an equivariant diffusion framework; a detailed comparison between this work and the work of Ref.~\cite{Aarts:2026zzr}  is presented in Section~\ref{subsec:comparison_aarts}.

In this work, we develop a diffusion model for two-dimensional SU(2) pure gauge theory.~The transition from U(1) to SU(2) introduces several conceptual and technical challenges.~First, the configuration space changes fundamentally:~U(1) link variables are phases on the circle $S^1$, while SU(2) link variables are elements of a three-dimensional manifold isomorphic to the 3-sphere $S^3$.~Second, the non-commutativity of SU(2) affects the structure of the theory, including the form of the action gradient.~We address these challenges through a quaternion parameterization that naturally respects the SU(2) manifold structure.~Notably, gauge equivariance is not built into the architecture but must be learned from data; while it seems natural to incorporate gauge invariance from the outset, as done in Ref.~\cite{Aarts:2026zzr}, it is instructive that a standard convolutional network can learn to approximate the correct gauge-field distributions without this constraint.~Moreover, since our uncorrected diffusion approach never evaluates the action during sampling, it may offer a starting point for future explorations of theories with a complex action, where Metropolis-based corrections are not directly applicable.

This paper is organized as follows.~Section~\ref{sec:su2_theory} reviews the mathematical framework of two-dimensional SU(2) lattice gauge theory.~Section~\ref{sec:diffusion} presents the diffusion model formalism adapted to gauge field configurations.~Section~\ref{sec:architecture} describes the neural network architecture and training procedure.~Section~\ref{sec:results} discusses our numerical results, including comparisons with exact analytical predictions and with the equivariant approach of Ref.~\cite{Aarts:2026zzr}.~Section~\ref{sec:Outlook} outlines future directions.

%=====================================================================
\section{SU(2) Lattice Gauge Theory: Mathematical Framework}
\label{sec:su2_theory}
%=====================================================================

We begin by establishing the mathematical foundations of two-dimensional SU(2) lattice gauge theory, which forms the physical basis for our diffusion model approach.

%---------------------------------------------------------------------
\subsection{The SU(2) Group and Quaternion Representation}
\label{subsec:su2_group}
%---------------------------------------------------------------------

The special unitary group SU(2) consists of all $2 \times 2$ complex matrices $U$ satisfying
\begin{equation}
    U^\dagger U = \mathbb{1}, \qquad \det(U) = 1.
    \label{eq:su2_definition}
\end{equation}
As a manifold, SU(2) is isomorphic to the 3-sphere $S^3 \subset \mathbb{R}^4$, which motivates the quaternion parameterization.~Every SU(2) element can be uniquely written as
\begin{equation}
    U = a_0 \mathbb{1} + i(a_1 \sigma_1 + a_2 \sigma_2 + a_3 \sigma_3),
    \label{eq:quaternion_rep}
\end{equation}
where $\sigma_i$ are the Pauli matrices
\begin{equation}
    \sigma_1 = \begin{pmatrix} 0 & 1 \\ 1 & 0 \end{pmatrix}, \quad
    \sigma_2 = \begin{pmatrix} 0 & -i \\ i & 0 \end{pmatrix}, \quad
    \sigma_3 = \begin{pmatrix} 1 & 0 \\ 0 & -1 \end{pmatrix},
    \label{eq:pauli_matrices}
\end{equation}
and the quaternion components $(a_0, a_1, a_2, a_3) \in \mathbb{R}^4$ satisfy the constraint
\begin{equation}
    a_0^2 + a_1^2 + a_2^2 + a_3^2 = 1.
    \label{eq:quaternion_constraint}
\end{equation}
This constraint defines the 3-sphere $S^3$ embedded in $\mathbb{R}^4$. In explicit matrix form, Eq.~\eqref{eq:quaternion_rep} becomes
\begin{equation}
    U = \begin{pmatrix} a_0 + i a_3 & a_2 + i a_1 \\ -a_2 + i a_1 & a_0 - i a_3 \end{pmatrix}.
    \label{eq:su2_matrix}
\end{equation}
The quaternion representation is particularly advantageous for numerical implementations:~group multiplication reduces to quaternion multiplication, the inverse is obtained by negating the vector part $(a_1, a_2, a_3) \to -(a_1, a_2, a_3)$, and the trace is simply $\Tr(U) = 2a_0$. Moreover, random sampling according to the Haar measure is achieved by drawing four independent Gaussian random numbers and normalizing to unit length, exploiting the uniformity of the Gaussian measure on $S^3$.

%---------------------------------------------------------------------
\subsection{Lattice Gauge Field Configuration}
\label{subsec:lattice_config}
%---------------------------------------------------------------------

On a two-dimensional Euclidean lattice $\Lambda$ of size $L_x \times L_t$ with periodic boundary conditions, a gauge field configuration consists of link variables $U_\mu(x) \in \text{SU}(2)$ associated with each oriented link from site $x$ to site $x + \hat{\mu}$, where $\hat{\mu}$ denotes the unit vector in direction $\mu \in \{0, 1\}$ (spatial and temporal directions, respectively).~The complete configuration is thus a collection
\begin{equation}
    \{U_\mu(x) : x \in \Lambda, \mu \in \{0, 1\}\},
    \label{eq:config_collection}
\end{equation}
containing $2 \times L_x \times L_t$ SU(2) matrices.~Using the quaternion representation, we store this as a tensor
\begin{equation}
    \mathcal{U} \in \mathbb{R}^{L_x \times L_t \times 2 \times 4},
    \label{eq:tensor_storage}
\end{equation}
where the last two indices correspond to direction $\mu$ and quaternion component $i$, respectively.

The fundamental gauge-invariant object is the plaquette, the elementary closed loop around a single lattice cell:
\begin{equation}
    P_{01}(x) = U_0(x) U_1(x + \hat{0}) U_0^\dagger(x + \hat{1}) U_1^\dagger(x),
    \label{eq:plaquette}
\end{equation}
where the product is taken around the elementary square in the counter-clockwise direction.~Since $P_{01}(x) \in \text{SU}(2)$, it can also be represented as a quaternion $(a_0^{(\text{plaq})}, a_1^{(\text{plaq})}, a_2^{(\text{plaq})}, a_3^{(\text{plaq})})$.

%---------------------------------------------------------------------
\subsection{Wilson Action and Observables}
\label{subsec:wilson_action}
%---------------------------------------------------------------------

The Wilson gauge action for two-dimensional SU(2) lattice gauge theory is defined as
\begin{equation}
    S[U] = -\frac{\beta}{2} \sum_{x \in \Lambda} \ReTr[P_{01}(x)] = -\beta \sum_{x \in \Lambda} a_0^{(\text{plaq})}(x),
    \label{eq:wilson_action}
\end{equation}
where $\beta = 4/g^2$ is the inverse gauge coupling (with $g$ the bare coupling constant), and $a_0^{(\text{plaq})}(x)$ is the $a_0$ component of the plaquette quaternion at site $x$.~The partition function is
\begin{equation}
    Z = \int \mathcal{D}U \, e^{-S[U]},
    \label{eq:partition_function}
\end{equation}
where the integration measure $\mathcal{D}U = \prod_{x,\mu} dU_\mu(x)$ is the product of Haar measures on SU(2).~Physical observables are computed as thermal averages
\begin{equation}
    \expect{O} = \frac{1}{Z} \int \mathcal{D}U \, O[U] \, e^{-S[U]}.
    \label{eq:thermal_average}
\end{equation}

The average plaquette is the most fundamental observable:
\begin{equation}
    \expect{P} = \left\langle \frac{1}{2} \ReTr[P_{01}] \right\rangle = \expect{a_0^{(\text{plaq})}}.
    \label{eq:average_plaquette}
\end{equation}
For two-dimensional SU(2) gauge theory, this has the exact analytical result~\cite{Creutz1980,Drouffe1983}
\begin{equation}
    \expect{P} = \frac{I_2(\beta)}{I_1(\beta)},
    \label{eq:exact_plaquette}
\end{equation}
where $I_n(\beta)$ denotes the modified Bessel function of the first kind.~Crucially, this result holds exactly on any finite periodic lattice, regardless of $L_x$ and $L_t$: in two dimensions with periodic boundary conditions, the single-plaquette integrals factorise in the partition function after gauge fixing to a maximal tree, so $\expect{P}$ receives no finite-volume corrections~\cite{Creutz1980,Drouffe1983}.~On a finite lattice, the average plaquette is computed as
\begin{equation}
    \expect{P} = \frac{1}{L_x L_t} \sum_{x \in \Lambda} \frac{1}{2} \ReTr[P_{01}(x)] = \frac{1}{L_x L_t} \sum_{x \in \Lambda} a_0^{(\text{plaq})}(x).
    \label{eq:finite_plaquette}
\end{equation}

%=====================================================================
\section{Diffusion Models: Theoretical Framework}
\label{sec:diffusion}
%=====================================================================

Diffusion models are a class of generative models that learn to reverse a gradual noising process.~In this section, we present the theoretical framework adapted to lattice gauge field configurations.

%---------------------------------------------------------------------
\subsection{Forward Diffusion Process}
\label{subsec:forward_diffusion}
%---------------------------------------------------------------------

Given an unnoised gauge field configuration represented as $\varphi_0 \in \mathbb{R}^{8 \times L_x \times L_t}$ (reshaped from the quaternion tensor, where 8 channels = 2 directions $\times$ 4 quaternion components), the forward process progressively adds Gaussian noise over $T$ timesteps according to the Markov chain
\begin{equation}
    q(\varphi_t | \varphi_{t-1}) = \mathcal{N}\left(\varphi_t; \sqrt{1 - \beta_t} \varphi_{t-1}, \beta_t \mathbb{1}\right),
    \label{eq:forward_step}
\end{equation}
where $\mathcal{N}(\varphi_t;\,{\mu},\,{\Sigma})$ denotes a Gaussian distribution over $\varphi_t$ with mean ${\mu}$ and covariance ${\Sigma}$, and $\{\beta_t\}_{t=1}^T$ is a variance schedule with $0 < \beta_t < 1$.

A key property of Gaussian diffusion is that $\varphi_t$ can be sampled directly from $\varphi_0$ without iterating through intermediate steps:
\begin{equation}
    q(\varphi_t | \varphi_0) = \mathcal{N}\left(\varphi_t; \sqrt{\bar{\alpha}_t} \varphi_0, (1 - \bar{\alpha}_t) \mathbb{1}\right),
    \label{eq:direct_sampling}
\end{equation}
where we define
\begin{equation}
    \alpha_t = 1 - \beta_t, \qquad \bar{\alpha}_t = \prod_{s=1}^t \alpha_s.
    \label{eq:alpha_definitions}
\end{equation}
This allows direct sampling via the reparameterization
\begin{equation}
    \varphi_t = \sqrt{\bar{\alpha}_t} \varphi_0 + \sqrt{1 - \bar{\alpha}_t} \epsilon, \quad \epsilon \sim \mathcal{N}(0, \mathbb{1}).
    \label{eq:reparameterization}
\end{equation}

We employ the cosine schedule~\cite{Nichol2021}:
\begin{equation}
    \bar{\alpha}_t = \cos^2\left(\frac{\pi}{2} \cdot \frac{t/T + s}{1 + s}\right),
    \label{eq:cosine_schedule}
\end{equation}
with small offset $s = 0.008$. This schedule ensures a smooth progression from $\bar{\alpha}_0 \approx 1$ (pure signal) to $\bar{\alpha}_T \approx 0$ (pure noise), avoiding the sharp transitions of linear schedules that can destabilize training.

%---------------------------------------------------------------------
\subsection{Reverse Diffusion Process}
\label{subsec:reverse_diffusion}
%---------------------------------------------------------------------

The reverse process learns to denoise by modeling the posterior $p_\theta(\varphi_{t-1} | \varphi_t)$. Following the denoising diffusion probabilistic models (DDPM) framework~\cite{Ho2020}, we parameterize this as a Gaussian
\begin{equation}
    p_\theta(\varphi_{t-1} | \varphi_t) = \mathcal{N}(\varphi_{t-1}; \mu_\theta(\varphi_t, t), \beta_t \mathbb{1}),
    \label{eq:reverse_posterior}
\end{equation}
where the mean is predicted by a neural network.~Rather than predicting $\mu_\theta$ directly, we train the network to predict the noise $\epsilon_\theta(\varphi_t, t)$, from which the mean is computed as
\begin{equation}
    \mu_\theta(\varphi_t, t) = \frac{1}{\sqrt{\alpha_t}} \left(\varphi_t - \frac{\beta_t}{\sqrt{1 - \bar{\alpha}_t}} \epsilon_\theta(\varphi_t, t)\right).
    \label{eq:mean_from_noise}
\end{equation}
The network is trained to minimize the denoising score matching objective:
\begin{equation}
    \mathcal{L} = \mathbb{E}_{t \sim \mathcal{U}(1,T), \varphi_0 \sim p_{\text{data}}, \epsilon \sim \mathcal{N}(0,\mathbb{1})} \left[\|\epsilon - \epsilon_\theta(\varphi_t, t)\|^2\right],
    \label{eq:score_matching_loss}
\end{equation}
where $\varphi_t$ is constructed from $\varphi_0$ and $\epsilon$ via Eq.~\eqref{eq:reparameterization}.

%---------------------------------------------------------------------
\subsection{Connection to Stochastic Quantization}
\label{subsec:stochastic_quantization}
%---------------------------------------------------------------------

Diffusion models have a deep connection to stochastic quantization~\cite{Parisi1981}, an alternative approach to quantizing field theories.~In stochastic quantization, a field $\varphi$ evolves in a fictitious Langevin time $\tau$ according to the equation
\begin{equation}
    \frac{\partial \varphi}{\partial \tau} = -\frac{\delta S[\varphi]}{\delta \varphi} + \sqrt{2} \eta(\tau),
    \label{eq:langevin}
\end{equation}
where $\eta$ is Gaussian white noise with $\expect{\eta(\tau) \eta(\tau')} = \delta(\tau - \tau')$. As $\tau \to \infty$, the distribution of $\varphi$ converges to the Boltzmann distribution $p(\varphi) \propto e^{-S[\varphi]}$.

The reverse diffusion process can be understood as a learned version of Langevin dynamics.~The score function, the gradient of the log-probability, is related to the action gradient:
\begin{equation}
    \nabla_\varphi \log p(\varphi) = -\nabla_\varphi S[\varphi].
    \label{eq:score_action_relation}
\end{equation}
The neural network $\epsilon_\theta$ learns an approximation to this score function, enabling sampling from the target distribution.~This connection provides a physics-based foundation for the diffusion process and suggests natural extensions to non-Abelian gauge theories through the established framework of stochastic quantization for gauge fields~\cite{Zwanziger1981}.~This connection between diffusion models and stochastic quantization in the context of lattice field theory was developed in Refs.~\cite{Wang:2023exq,Wang:2023sry} and exploited for gauge field simulations in Refs.~\cite{Zhu2024a,Zhu2024b}.

%---------------------------------------------------------------------
\subsection{Physics-Conditioned Sampling}
\label{subsec:physics_conditioning}
%---------------------------------------------------------------------

A key advantage of our approach is the ability to generate configurations at coupling values $\beta$ different from the training coupling $\beta_0$, without retraining the model.~Theoretically, this follows from the structure of the Wilson action and the connection to stochastic quantization.

Since $S[U] = \beta \tilde{S}[U]$ where $\tilde{S} = -\sum_x a_0^{(\text{plaq})}(x)$ is independent of $\beta$, the score function scales linearly with the coupling:
\begin{equation}
    \nabla_\varphi \log p^{(\beta)}(\varphi) = -\beta \nabla_\varphi \tilde{S}[\varphi].
    \label{eq:score_scaling}
\end{equation}
A model trained at $\beta_0$ learns the score $s^{(\beta_0)} \propto -\beta_0 \nabla \tilde{S}$. To generate configurations at coupling $\beta$, we rescale the predicted noise\footnote{This physics-conditioned approach was introduced in Refs.~\cite{Zhu2024a,Zhu2024b}, building on the framework of Refs.~\cite{Wang:2023exq,Wang:2023sry}.}:
\begin{equation}
    \hat{\epsilon}^{(\beta)} = \frac{\beta}{\beta_0} \hat{\epsilon}^{(\beta_0)}.
    \label{eq:noise_rescaling}
\end{equation}
In practice, when extrapolating to couplings far from $\beta_0$, we stabilise the sampling by interpolating between a constant baseline and the rescaled network prediction:
\begin{equation}
    \hat{\epsilon}^{(\beta)} = (1-\lambda) + \lambda \cdot \frac{\beta}{\beta_0}\hat{\epsilon}^{(\beta_0)},
    \label{eq:blended_noise}
\end{equation}
where $\lambda \in [0.15,\, 0.5]$ controls the network contribution.~Results are robust across this range, with smaller values preferred for $\beta \gg \beta_0$, where the sampling initialisation is also adapted accordingly.~This allows exploration of various coupling regions without generating new training data, dramatically reducing computational cost.
%=====================================================================
\section{Neural Network Architecture and Implementation}
\label{sec:architecture}
%=====================================================================

%---------------------------------------------------------------------
\subsection{Data Representation}
\label{subsec:data_representation}
%---------------------------------------------------------------------

The gauge field configuration tensor $\mathcal{U} \in \mathbb{R}^{L_x \times L_t \times 2 \times 4}$ is reshaped to $\varphi \in \mathbb{R}^{8 \times L_x \times L_t}$, where the 8 channels correspond to the $2 \times 4 = 8$ quaternion components (2 directions $\times$ 4 components each \footnote{The channel ordering is:~ $\varphi = \left[\vec{a}^{(\mu=0)},  \vec{a}^{(\mu=1)} \right]$.}).~This representation treats the gauge field as an ``8-channel image'' over the $L_x \times L_t$ lattice, enabling the use of standard convolutional architectures while preserving the full gauge field information.

%---------------------------------------------------------------------
\subsection{Training Data Generation and Gauge Augmentation}
\label{subsec:data_generation}
%---------------------------------------------------------------------

Training configurations are generated using the Hybrid Monte Carlo algorithm~\cite{Duane1987} at a fixed coupling $\beta_0$. The HMC algorithm combines molecular dynamics evolution with a Metropolis accept/reject step, providing exact sampling from the Boltzmann distribution while maintaining reasonable acceptance rates.

The gauge symmetry of lattice gauge theory provides a powerful mechanism for data augmentation.~Given a configuration $U$, we generate additional training samples by applying random gauge transformations:
\begin{equation}
    U'_\mu(x) = \Omega(x) U_\mu(x) \Omega^\dagger(x + \hat{\mu}),
    \label{eq:gauge_transform}
\end{equation}
where $\{\Omega(x)\}$ are independent random SU(2) elements sampled according to the Haar measure.~In practice, we generate Haar-random SU(2) elements by sampling 4 independent Gaussian random numbers and normalizing to unit quaternion:
\begin{equation}
    \Omega \leftrightarrow \frac{(g_0, g_1, g_2, g_3)}{\sqrt{g_0^2 + g_1^2 + g_2^2 + g_3^2}}, \qquad g_i \sim \mathcal{N}(0, 1).
    \label{eq:haar_sampling}
\end{equation}

All gauge-transformed configurations share identical physical content, the same action, plaquette values, Wilson loops, and any gauge-invariant observable, but have different numerical values for the link variables.~This augmentation teaches the network the gauge equivalence of configurations and improves generalization.~For each of the 10,000 HMC configurations, we generate one gauge-transformed copy, yielding a total of 20,000 training samples.
%---------------------------------------------------------------------
\subsection{U-Net Architecture}
\label{subsec:unet}
%---------------------------------------------------------------------

\begin{figure}[htbp]
    \centering
    \includegraphics[width=0.6\textwidth]{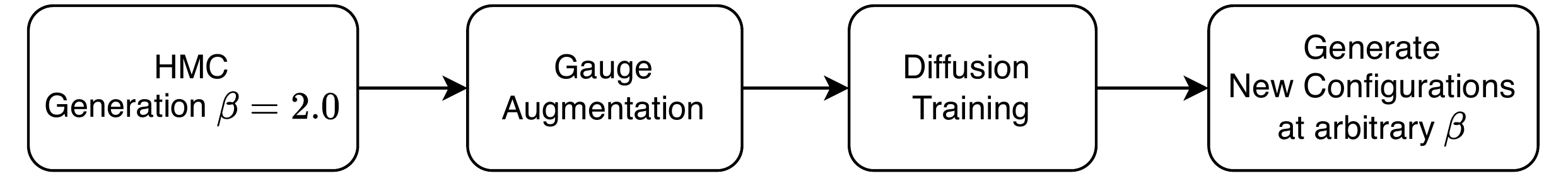}
    \caption{Complete pipeline of the diffusion model approach. Training data is generated via HMC at coupling $\beta_0$, augmented through random gauge transformations, and used to train the diffusion model. Once trained, the model generates new configurations at arbitrary coupling $\beta$ and lattice extents without retraining.}
    \label{fig:overall_structure}
\end{figure}

The noise prediction network $\epsilon_\theta(\varphi_t, t)$ is implemented as a U-Net architecture~\cite{Ronneberger2015} with specific adaptations for lattice gauge fields.~The overall structure of our diffusion model pipeline is illustrated in Fig.~\ref{fig:overall_structure}.

\begin{figure}[htbp]
    \centering
    \includegraphics[width=0.8\textwidth]{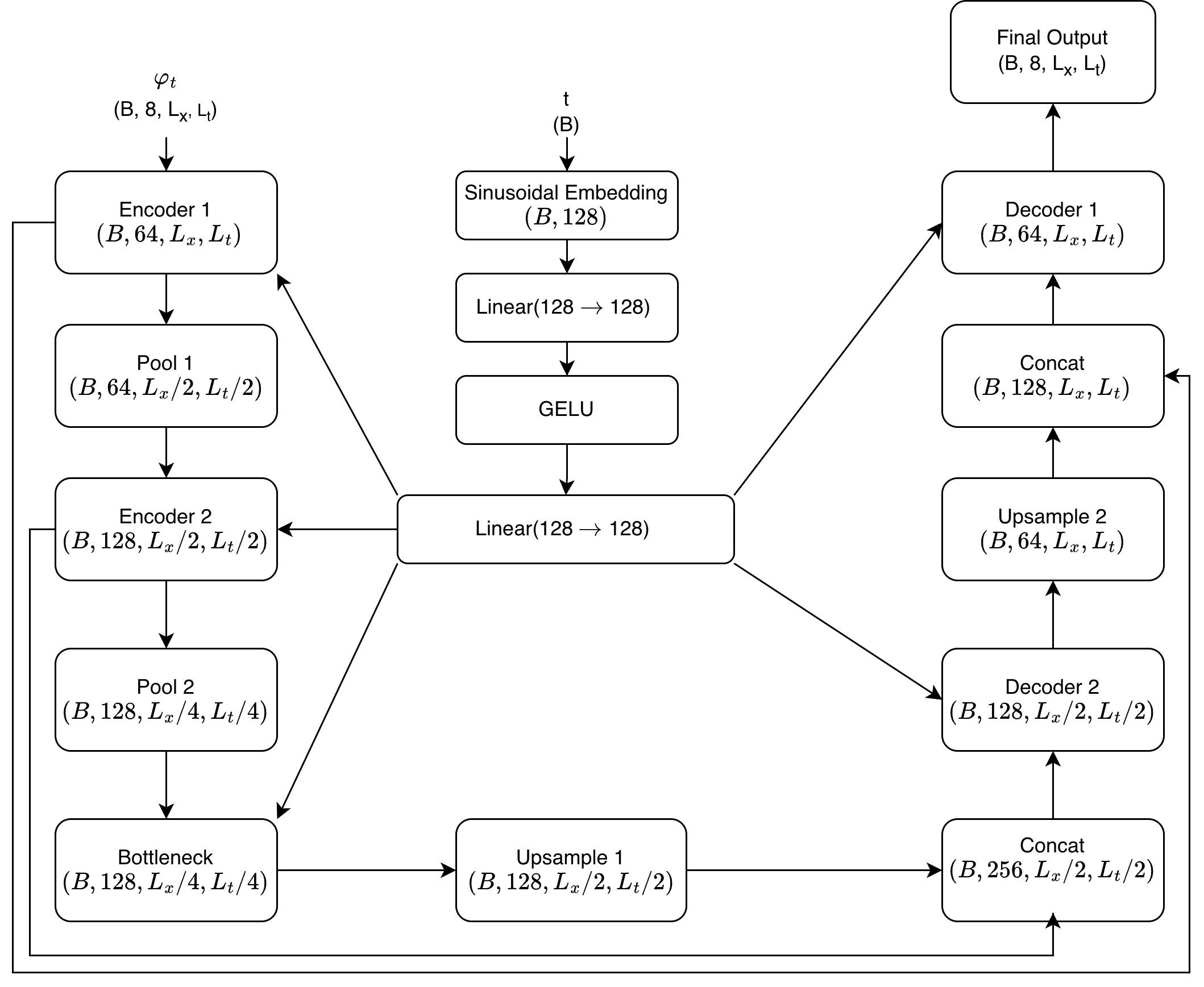}
    \caption{Detailed architecture of the U-Net neural network $\epsilon_\theta$. The network takes the noised configuration $\varphi_t$ and timestep $t$ as inputs and predicts the noise $\epsilon_{\text{pred}}$.~The encoder-decoder structure with skip connections enables multi-scale feature extraction while maintaining spatial resolution in the output.}
    \label{fig:unet_architecture}
\end{figure}

The network consists of an encoder path that progressively downsamples the spatial dimensions while increasing feature channels, a bottleneck, and a decoder path that upsamples back to the original resolution with skip connections from the encoder.~The architecture is detailed in Fig.~\ref{fig:unet_architecture}.

\subsubsection{Periodic Boundary Conditions}

Lattice gauge theory is conventionally formulated on a torus with periodic boundary conditions.~Standard neural network padding schemes (zero padding, reflection padding) would introduce artificial boundary effects and break translational invariance.~We instead use \textit{circular padding}, which wraps the input around so that sites at opposite boundaries are correctly identified as neighbors.

For a convolutional kernel of size $k$ with padding $p = \lfloor k/2 \rfloor$, circular padding extends the input by copying values from the opposite boundary:
\begin{equation}
    \varphi_{\text{padded}}[i] = \varphi[(i - p) \mod L],
    \label{eq:circular_padding}
\end{equation}
where $L$ is the lattice extent.~This ensures translational invariance is exactly preserved by the convolution operation, which is essential for the fully convolutional architecture to generalize to different lattice sizes.

\subsubsection{Time Embedding}

The diffusion timestep $t$ is encoded using sinusoidal positional embeddings~\cite{Vaswani2017}, which map the scalar timestep to a high-dimensional vector:
\begin{equation}
    \text{PE}^{2i}(t) = \sin\left(\frac{t}{10^{6i/d}}\right), \quad \text{PE}^{2i+1}(t) = \cos\left(\frac{t}{10^{6i/d}}\right),
    \label{eq:time_embedding}
\end{equation}
where $d$ is the embedding dimension (we use $d = 128$).~This embedding is processed through a Multi-Layer Perceptron (MLP) with the Gaussian Error Linear Unit (GELU) activations, and added to the feature maps in each residual block, allowing the network to condition its behavior on the noise level.

\subsubsection{Residual Blocks}

Each encoder and decoder level consists of convolutional residual blocks with the following structure:
\begin{enumerate}
    \item GroupNorm $\to$ GELU activation $\to$ Conv2d (circular padding)
    \item Add time embedding (projected to appropriate channel dimension)
    \item GroupNorm $\to$ GELU activation $\to$ Conv2d (circular padding)
    \item Residual connection (with $1 \times 1$ convolution if channel dimensions differ)
\end{enumerate}
GroupNorm~\cite{Wu2018} is preferred over BatchNorm for the small batch sizes typical in physics applications, as it normalizes over channel groups rather than requiring large batch statistics.

%---------------------------------------------------------------------
\subsection{Training Procedure}
\label{subsec:training}
%---------------------------------------------------------------------

The training procedure follows the standard DDPM approach, as illustrated in Fig.~\ref{fig:training_step}. For each batch of clean configurations $\varphi_0$:
\begin{enumerate}
    \item Sample random timesteps $t^i \sim \mathcal{U}(1, T)$ independently for each configuration $i$ in the batch;
    \item Sample random noise $\epsilon^i_{\text{true}} \sim \mathcal{N}(0, \mathbb{1})$;
    \item Construct noised configurations $\varphi_t^i = \sqrt{\bar{\alpha}_{t^i}} \varphi_0^i + \sqrt{1 - \bar{\alpha}_{t^i}} \epsilon^i_{\text{true}}$;
    \item Forward pass through the network to predict $\epsilon^i_{\text{pred}} = \epsilon_\theta(\varphi_t^i, t^i)$;
    \item Compute MSE loss $\mathcal{L} = \frac{1}{B}\sum_i \|\epsilon^i_{\text{true}} - \epsilon^i_{\text{pred}}\|^2$;
    \item Backpropagate and update network parameters.
\end{enumerate}

We use the Adam optimizer~\cite{Kingma2015} with learning rate $10^{-4}$ and cosine annealing schedule.~Training is performed for 100 epochs with batch size 32.~Gradient clipping with maximum norm 1.0 prevents training instabilities.

%---------------------------------------------------------------------
\subsection{Sampling Algorithm}
\label{subsec:sampling}
%---------------------------------------------------------------------

\begin{figure}[htbp]
    \centering
    \begin{subfigure}[t]{0.48\textwidth}
        \centering
        \includegraphics[width=\textwidth]{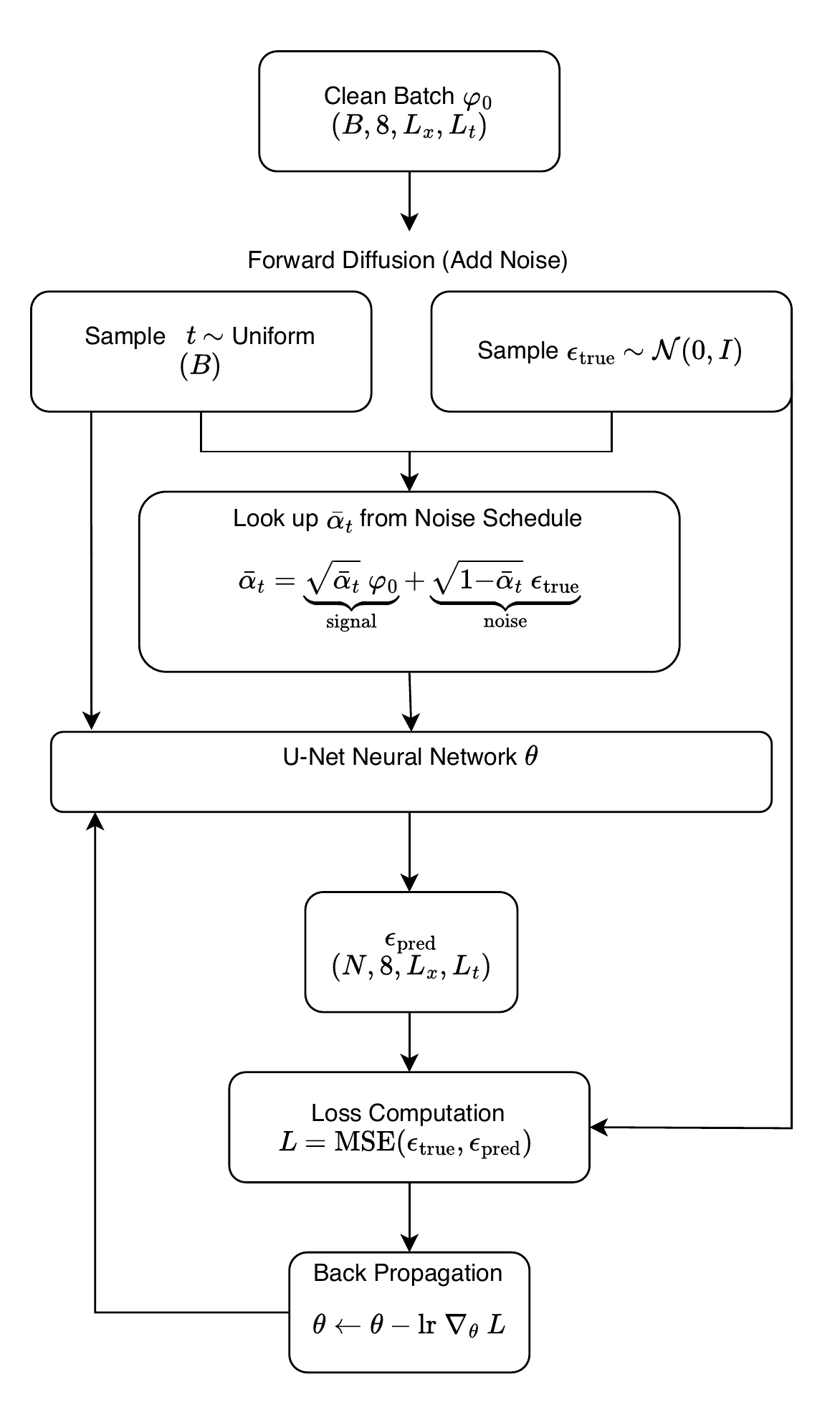}
        \caption{}
        \label{fig:training_step}
    \end{subfigure}
    \hfill
    \begin{subfigure}[t]{0.44\textwidth}
        \centering
        \includegraphics[width=\textwidth]{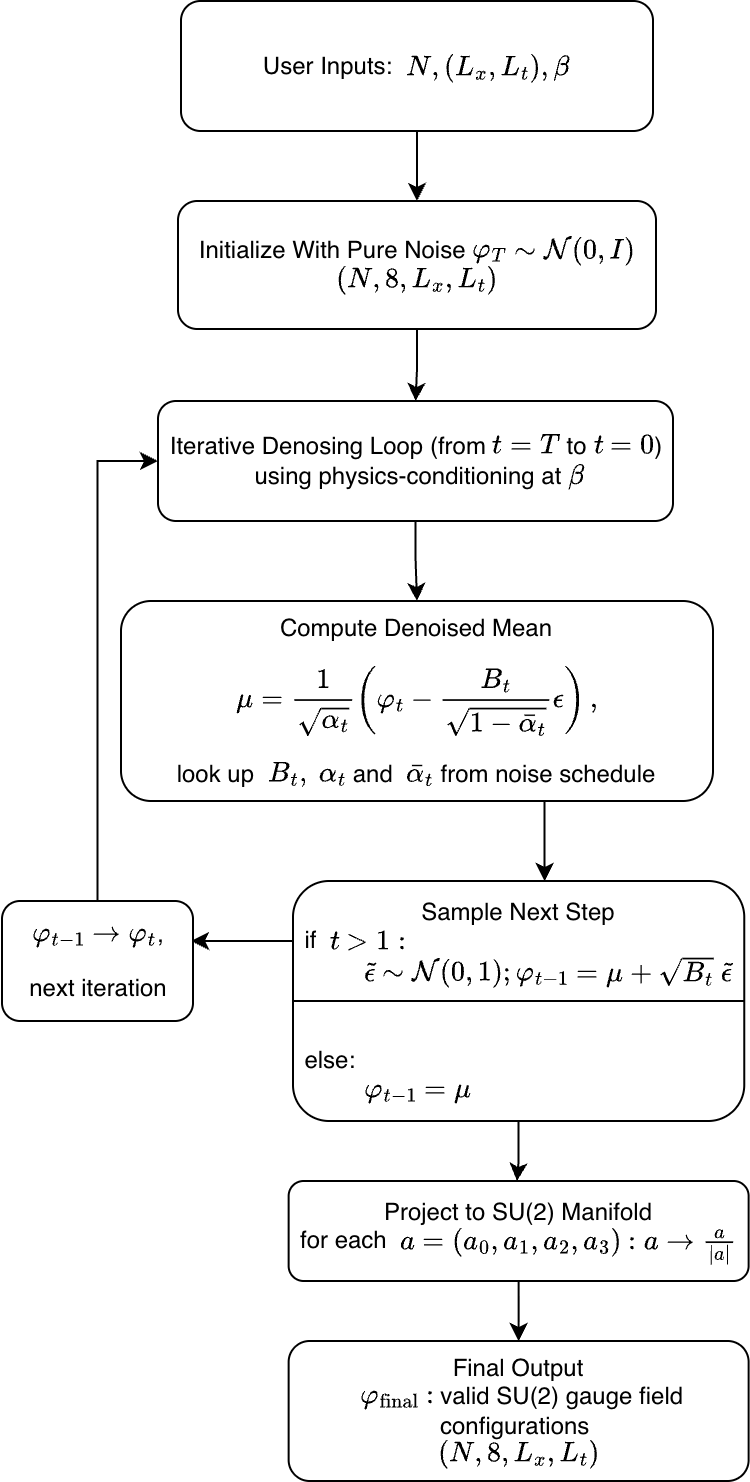}
        \caption{}
        \label{fig:sampling}
    \end{subfigure}
    \caption{Overview of the training and sampling procedures.~(a)~Flowchart of a single training step.~The model learns to predict the noise $\epsilon_{\text{true}}$ that was added to clean configurations, enabling the reverse denoising process during sampling.~(b)~Sampling algorithm flowchart.~Starting from Gaussian noise, the iterative denoising loop applies physics-conditioned noise prediction and gradually recovers a gauge field configuration.~The final projection step ensures the output satisfies the SU(2) constraint.}
    \label{fig:training_and_sampling}
\end{figure}

To generate new configurations, we reverse the diffusion process starting from pure noise, as illustrated in Fig.~\ref{fig:sampling}. 

The final quaternion normalization step projects the generated configuration back onto the SU(2) manifold, correcting any small deviations from unit norm accumulated during sampling.~This projection is essential for ensuring the output represents valid SU(2) gauge links.

%=====================================================================
\section{Results}
\label{sec:results}
%=====================================================================

We now present our numerical results, investigating the model's ability to generate configurations that reproduce exact analytical predictions across a range of couplings and lattice sizes.

\subsection{Validation at Training Coupling}
\label{subsec:validation_training}
%---------------------------------------------------------------------

We first validate the model at the training coupling $\beta_0 = 2.0$ on the training geometry $8 \times 8$.~Generating 500 configurations, we compute the average plaquette and compare with the exact analytical prediction Eq.~\eqref{eq:exact_plaquette}.

The diffusion model yields $\expect{P}_{\text{diff}} = 0.4329 \pm 0.0001\,(\text{stat})$, where the statistical uncertainty $\sigma_{\text{stat}}$ is the autocorrelation-corrected standard error of the mean from 500 configurations.~The exact value is $\expect{P}_{\text{exact}} = I_2(2)/I_1(2) = 0.4331$, giving a bias $\Delta \equiv \expect{P}_{\text{diff}} - \expect{P}_{\text{exact}} = -0.0002$, well within statistical precision.

Throughout this work, since the exact answer is known analytically, we report both $\sigma_{\text{stat}}$ and the bias $\Delta$ for each $(\beta, L_x \times L_t)$ point as the most transparent diagnostics of model quality.~As established in Eq.~\eqref{eq:exact_plaquette}, the exact plaquette $I_2(\beta)/I_1(\beta)$ holds on any finite periodic lattice, independent of $L_x$ and $L_t$: in two dimensions the single-plaquette expectation values factorise after gauge-fixing to a maximal tree~\cite{Creutz1980,Drouffe1983}.~Consequently, a nonzero $\Delta$ at any lattice geometry is a direct measure of model error.

\begin{figure}[htbp]
    \centering
    \begin{subfigure}[t]{0.48\textwidth}
        \centering
        \includegraphics[width=\textwidth]{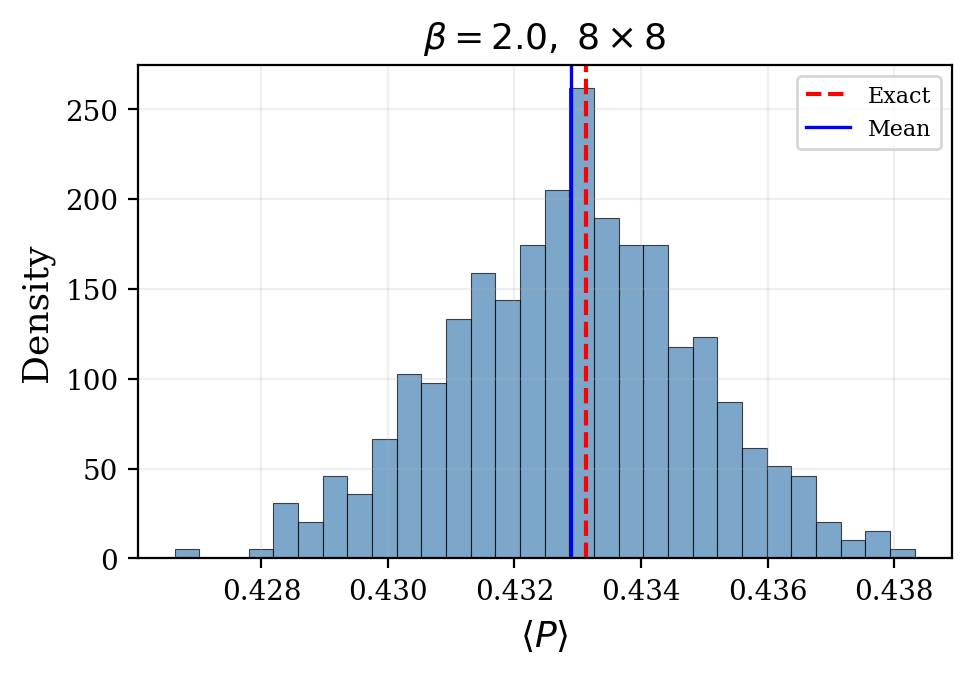}
        \caption{}
        \label{fig:plaquette_dist_8x8}
    \end{subfigure}
    \hfill
    \begin{subfigure}[t]{0.48\textwidth}
        \centering
        \includegraphics[width=\textwidth]{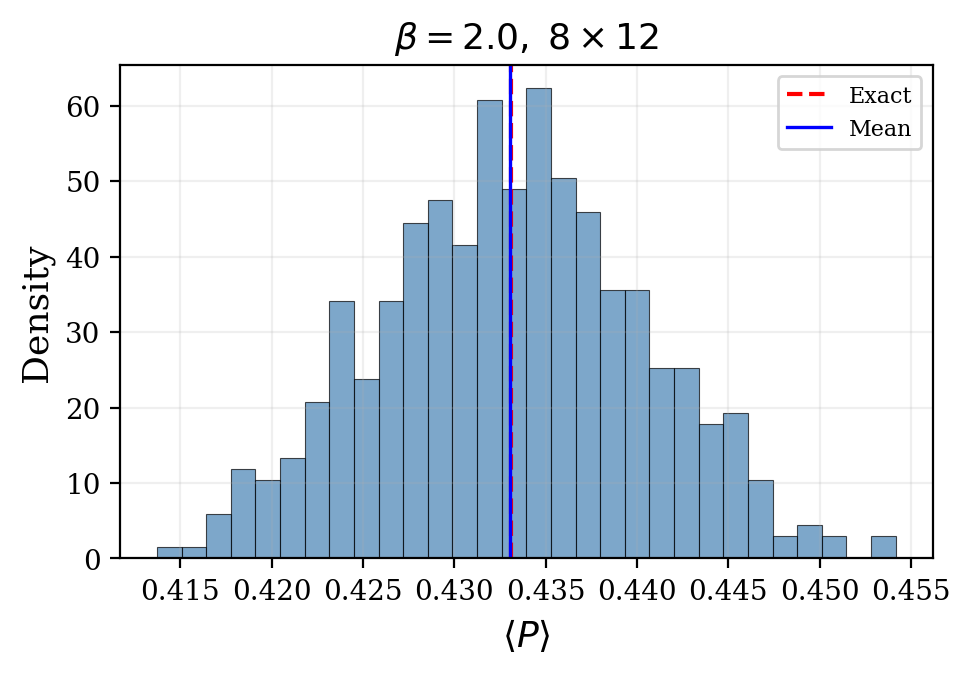}
        \caption{}
        \label{fig:plaquette_dist_8x12}
    \end{subfigure}

    \vspace{0.5em}

    \begin{subfigure}[t]{0.48\textwidth}
        \centering
        \includegraphics[width=\textwidth]{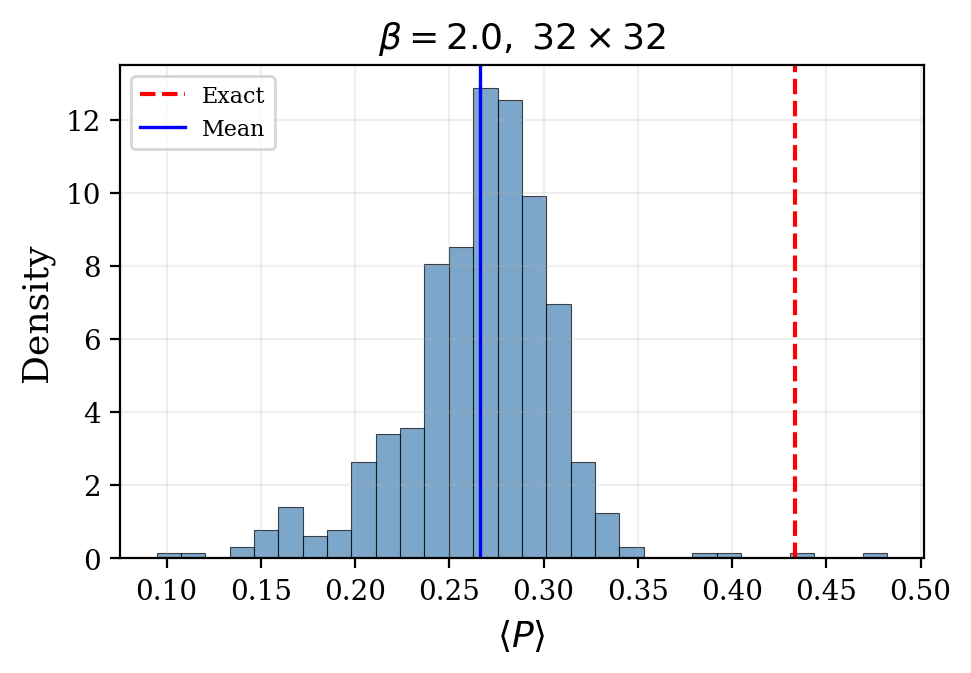}
        \caption{}
        \label{fig:plaquette_dist_32x32}
    \end{subfigure}
    \caption{Plaquette distribution from 500 diffusion-generated configurations at $\beta = 2.0$.~Red dashed line: exact value $I_2(\beta)/I_1(\beta)$; blue line: sample mean.~(a)~$8 \times 8$ training lattice.~The distribution is well-centered on the exact result.~(b)~$8 \times 12$ lattice.~The distribution remains centered near the exact result, confirming successful generalisation to geometries slightly larger than the training lattice.~(c)~$32 \times 32$ lattice.~The distribution is visibly shifted from the exact value, reflecting the difficulty of generalising to lattice sizes much larger than the $8 \times 8$ training geometry.}
    \label{fig:plaquette_dist_all}
\end{figure}

Figure~\ref{fig:plaquette_dist_8x8} shows the plaquette distribution for the training point. The distribution is well-centered on the exact value, with the sample mean nearly coinciding with it, confirming that the model has learned the correct equilibrium distribution at $(\beta_0, 8\times 8)$.

%---------------------------------------------------------------------
\subsection{Physics-Conditioned Generation at Different Couplings}
\label{subsec:beta_generalization}
%---------------------------------------------------------------------

A central capability of our approach is generating configurations at couplings different from the training value.~Using physics-conditioned sampling (Section~\ref{subsec:physics_conditioning}), we generate 500 configurations for $\beta \in \{1.0, 1.5, 2.0, 2.5, 3.0, 3.5, 4.0\}$ on the \emph{training geometry} $8 \times 8$ and compare with exact predictions.

\begin{figure}[htbp]
    \centering
    \begin{subfigure}[t]{0.48\textwidth}
        \centering
        \includegraphics[width=\textwidth]{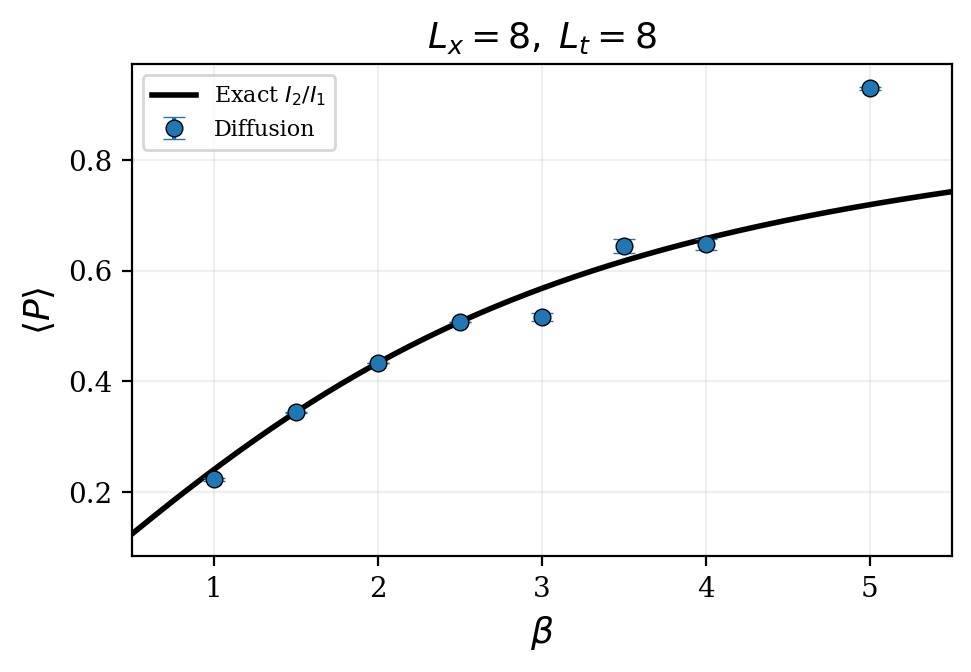}
        \caption{}
        \label{fig:plaquette_vs_beta_8x8}
    \end{subfigure}
    \hfill
    \begin{subfigure}[t]{0.48\textwidth}
        \centering
        \includegraphics[width=\textwidth]{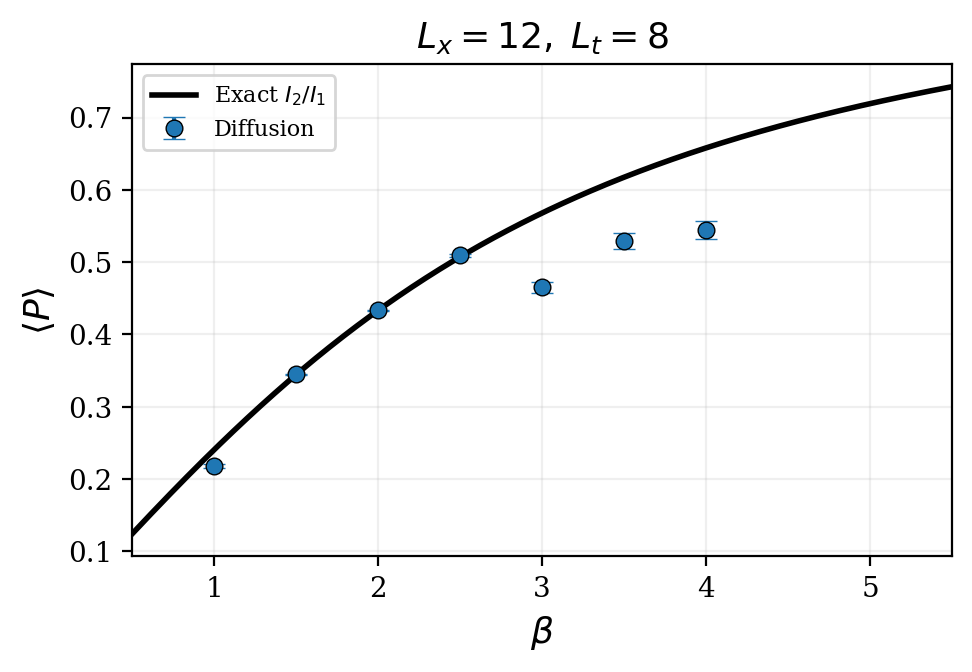}
        \caption{}
        \label{fig:plaquette_vs_beta}
    \end{subfigure}

    \vspace{0.5em}

    \begin{subfigure}[t]{0.48\textwidth}
        \centering
        \includegraphics[width=\textwidth]{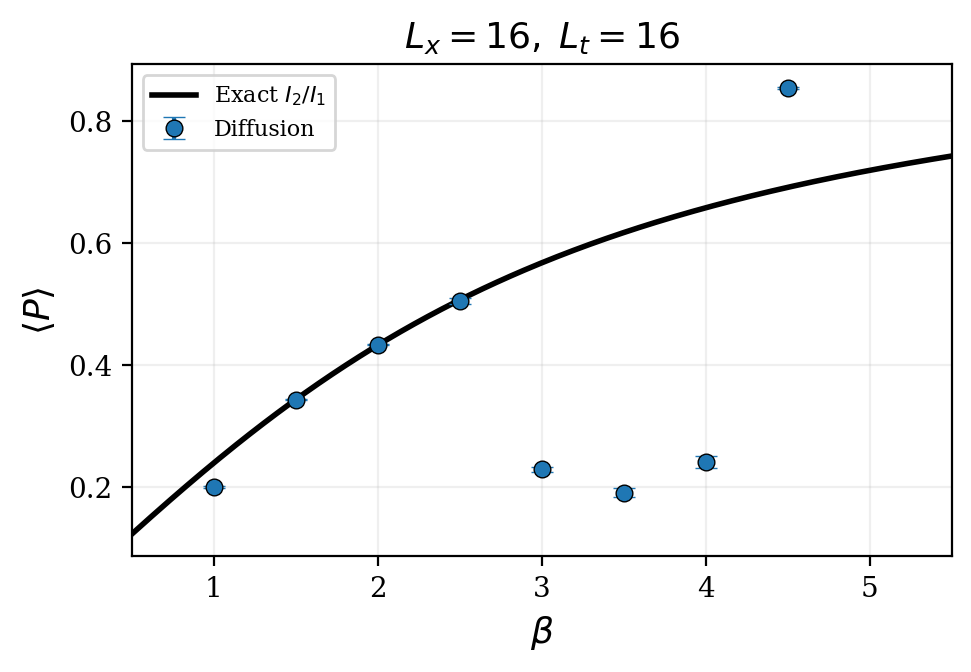}
        \caption{}
        \label{fig:plaquette_vs_beta_16x16}
    \end{subfigure}
    \caption{Average plaquette $\expect{P}$ as a function of coupling $\beta$.~Blue circles:~diffusion model results with $\sigma_{\text{stat}}$ error bars; solid line:~exact analytical result $I_2(\beta)/I_1(\beta)$.~(a)~$8 \times 8$ training lattice.~Biases are $|\Delta| \leq 0.001$ near the training coupling and remain below $0.06$ across $\beta \in [1, 4]$.~(b)~$12 \times 8$ lattice, which shares the training extent $L=8$.~The model achieves good agreement across the range $\beta \in [1, 4]$.~(c)~$16 \times 16$ lattice.~The model achieves near-exact agreement for $\beta \in [1.5, 2.5]$.}
    \label{fig:plaquette_vs_beta_all}
\end{figure}

Figure~\ref{fig:plaquette_vs_beta_8x8} shows the average plaquette as a function of $\beta$.~Near the training coupling, the model achieves excellent agreement: at $\beta = 1.5$, $2.0$, and $2.5$ the biases are $|\Delta| \leq 0.0004$.~At the boundaries of the explored coupling range, larger biases appear: $|\Delta| = 0.017$ at $\beta = 1.0$ and $|\Delta| = 0.051$ at $\beta = 3.0$, with intermediate values $|\Delta| \approx 0.01$--$0.03$ at $\beta = 3.5$ and $4.0$.~This demonstrates that physics-conditioned rescaling captures the coupling dependence of the score function well near the training point, with accuracy degrading gradually at larger $|\beta - \beta_0|$.

\begin{table}[htbp]
\centering
\caption{Average plaquette at $\beta = 2.0$ for increasing lattice sizes.  $\sigma_{\text{stat}}$ is the autocorrelation-corrected statistical error; $\Delta = \expect{P}_{\text{diff}} - \expect{P}_{\text{exact}}$ is the bias.  The exact result $I_2(\beta)/I_1(\beta) = 0.4331$ is volume-independent (see text).}
\label{tab:plaquette_volume}
\begin{tabular}{ccccc}
\hline\hline
$L_x \times L_t$ & $\expect{P}_{\text{diff}}$ & $\sigma_{\text{stat}}$ & $\Delta$ & $|\Delta|/\sigma_{\text{stat}}$ \\
\hline
$8 \times 8$   & 0.4329 & 0.0001 & $-0.0002$ & $2.7$ \\
$8 \times 12$  & 0.4331 & 0.0003 & $-0.0001$ & $0.2$ \\
$12 \times 8$  & 0.4333 & 0.0002 & $+0.0001$ & $0.6$ \\
$12 \times 12$ & 0.3892 & 0.0041 & $-0.0439$ & $10.8$ \\
$16 \times 12$ & 0.3594 & 0.0037 & $-0.0738$ & $20.1$ \\
$16 \times 16$ & 0.4339 & 0.0008 & $+0.0008$ & $0.9$ \\
$32 \times 32$ & 0.2661 & 0.0015 & $-0.1670$ & $108$ \\
\hline\hline
\end{tabular}
\end{table}

%---------------------------------------------------------------------
\subsection{Lattice Size Generalisation}
\label{subsec:size_generalization}
%---------------------------------------------------------------------

The fully convolutional architecture with periodic padding enables generation on lattices of different sizes without retraining.~Since $\expect{P}$ is exactly volume-independent in two dimensions (Sec.~\ref{subsec:wilson_action}), the bias $\Delta$ on a non-training geometry directly measures how well the model has learned to reproduce the correct local gauge-field structure beyond its training domain.~We test this by generating configurations on lattices with extents $L_x, L_t \in \{8, 12, 16, 32\}$, all using the model trained on $8 \times 8$ configurations, across the full range of couplings $\beta \in \{1.0, 1.5, 2.0, 2.5, 3.0, 3.5, 4.0\}$.

A clear hierarchy emerges, governed by how close the lattice geometry is to the training size:
\begin{itemize}
 \item $[L_x, L_t] = [8 \times 8]$:~biases $|\Delta| \leq 0.001$ for $\beta \in [1.5, 2.5]$, growing to $|\Delta| \approx 0.02$--$0.05$ at the extremes of the coupling range.
 \item $[L_x, L_t] \in \{ [8 \times 12], [12 \times 8], [8 \times 16], [16 \times 8]\}$:~biases $|\Delta| \lesssim 0.003$ for $\beta \in [1.5, 2.5]$, growing significantly for $\beta \geq 3$ and also elevated at $\beta = 1.0$ ($|\Delta| \approx 0.02$).
 \item $[12 \times 12], [12 \times 16], [16 \times 12]$:~moderate biases $|\Delta| \sim 0.02$--$0.07$ at low $\beta$, reaching $|\Delta| > 0.2$ at higher couplings.
 \item $[16 \times 16]$:~near-exact agreement ($|\Delta| < 0.002$) for $\beta \in [1.5, 2.5]$, with large deviations at higher couplings and at $\beta = 1.0$ ($|\Delta| \approx 0.04$).
 \item $[32 \times 32]$:~significant biases $|\Delta| > 0.15$ at most couplings.
\end{itemize}

Figures~\ref{fig:plaquette_vs_beta}--\ref{fig:plaquette_vs_beta_16x16} illustrate this progression.~For the $12 \times 8$ lattice (Fig.~\ref{fig:plaquette_vs_beta}), which shares the training extent $L=8$ in one direction, the model tracks the exact curve well up to $\beta \approx 3$.~For the $16 \times 16$ lattice (Fig.~\ref{fig:plaquette_vs_beta_16x16}), near-exact agreement is achieved for $\beta \in [1.5, 2.5]$, with deviations appearing at higher couplings and at $\beta = 1.0$.

The plaquette distributions further illustrate the volume dependence.~Figure~\ref{fig:plaquette_dist_8x12} shows the $8 \times 12$ distribution at $\beta = 2.0$, which remains well-centered on the exact value. In contrast, Fig.~\ref{fig:plaquette_dist_32x32} for the $32 \times 32$ lattice shows a visible shift, with the distribution peaked well below the exact result.

This behaviour is expected: the model learns local correlations at the scale of the training lattice, and larger volumes require the network to extrapolate beyond its training domain. Nevertheless, the successful generalisation to lattices up to $16 \times 16$, four times the training volume, demonstrates the utility of the fully convolutional approach.

\subsection{HMC autocorrelation analysis}
\label{sec:hmc_autocorrelation}

\begin{figure}[htbp]
    \centering
    \includegraphics[width=0.5\textwidth]{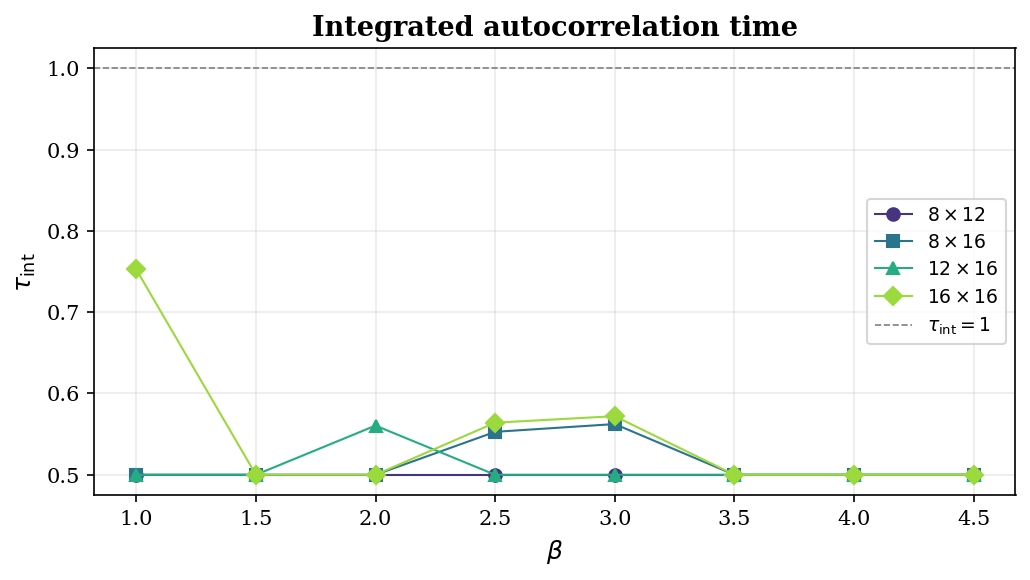}
    \caption{Integrated autocorrelation time $\tau_{\mathrm{int}} = \tfrac{1}{2} + \sum_{t=1} R(t)$ of the average plaquette as a function of $\beta$ for all lattice sizes.~All values remain near the theoretical minimum $\tau_{\mathrm{int}}=0.5$, indicating that consecutive configurations are effectively uncorrelated.}
    \label{fig:hmc_autocorr}
\end{figure}

To verify that the HMC-generated configurations are statistically independent, we computed the normalised autocorrelation function $R(t)$ of the average plaquette for each $(\beta, L_x, L_t)$ point and extracted the integrated autocorrelation time $\tau_{\mathrm{int}}$.
Figure~\ref{fig:hmc_autocorr} shows $\tau_{\mathrm{int}}$ as a function of $\beta$ for all lattice sizes.
Across the entire parameter grid $\tau_{\mathrm{int}} < 1$, confirming that consecutive HMC trajectories produce effectively independent configurations and no thinning is required.

\subsection{Comparison with Equivariant Diffusion Models}
\label{subsec:comparison_aarts}
%---------------------------------------------------------------------

During the preparation of this manuscript, Aarts~\textit{et al.}~\cite{Aarts:2026zzr} presented a gauge equivariant diffusion framework for two-dimensional U(2) and SU(2) gauge theories.~Since both works address the same physical system with diffusion models trained at $\beta_0 = 2$, a detailed comparison is instructive.

The two approaches differ in three fundamental respects.~First, the treatment of gauge symmetry: Ref.~\cite{Aarts:2026zzr} formulates the diffusion process directly on the group manifold via a Stratonovich SDE (Eq.~(4) therein) and employs lattice gauge equivariant convolutional neural networks (L-CNNs)~\cite{Favoni:2020reg} as the score network (Sec.~``Gauge equivariant diffusion models'' therein), thereby enforcing exact local gauge equivariance at the architectural level through the transformation law of Eq.~(6) therein.~Our approach instead works in the flat quaternion representation $(a_0, a_1, a_2, a_3)$ of SU(2) and uses a standard convolutional U-Net, relying on the unit-quaternion projection to return configurations to the group manifold.~Gauge equivariance is not enforced but must be learned from data.

Second, the sampling algorithm:~Ref.~\cite{Aarts:2026zzr} employs the Metropolis-adjusted annealed Langevin algorithm (MAALA, Eqs.~(11)--(12) therein), which applies a Metropolis accept/reject step based on the target Wilson action during the final $2\%$ of diffusion time steps (End Matter, ``MAALA implementation details'' therein).~This correction guarantees that generated samples are asymptotically distributed according to $\exp(-S_E)$ at the target $\beta$, regardless of residual model errors.~Our approach uses uncorrected reverse diffusion with physics-conditioned noise rescaling (Section~\ref{subsec:physics_conditioning}), so the quality of the generated ensemble depends entirely on the accuracy of the learned score function.

Third, the training setup differs in scale:~Ref.~\cite{Aarts:2026zzr} trains on $10^5$ HMC configurations at $\beta_0 = 2$ on a $16 \times 16$ lattice using an ensemble of ten independently initialised networks with ${\sim}\,15\,000$ parameters each (Sec.~``Numerical results'' and ``Training and model details'' therein), while our model is trained on a smaller set at $\beta_0 = 2$ on an $8 \times 8$ lattice with a single network.

Despite these differences, qualitative similarities emerge.~Both models successfully generalise to couplings away from the training point using score rescaling: Ref.~\cite{Aarts:2026zzr} reports accurate Wilson loop predictions up to $\beta \approx 14$ on $L = 16$ (Fig.~3 therein), while our model achieves biases $|\Delta| \leq 0.001$ for $\beta \in [1.5, 2.5]$ on the $8 \times 8$ training geometry, growing to $|\Delta| \approx 0.05$ at $\beta = 3$.~Both also demonstrate volume generalisation through fully convolutional architectures.

The key quantitative differences can be attributed to the Metropolis correction.~The MAALA acceptance rates reported in Ref.~\cite{Aarts:2026zzr} (Fig.~4 therein) remain above $50\%$ for $\beta \lesssim 12$ at $L = 16$, ensuring that even where the score network becomes inaccurate, the generated samples are corrected towards the true distribution.~Our uncorrected approach has no such safety net:~deviations from the exact plaquette grow with both $\beta$ and lattice volume for non-training geometries (Table~\ref{tab:plaquette_volume}), reaching $|\Delta| = 0.17$ at $32 \times 32$.~For the SU(2) plaquette at $\beta = 2$ on $L = 16$, Ref.~\cite{Aarts:2026zzr} reports $\expect{P} = 0.4333(2)$ (Table~II therein) against the exact value $0.4331$, while our model on the same-volume $16 \times 16$ lattice yields $\expect{P} = 0.4339 \pm 0.0008\,(\text{stat})$, with bias $\Delta = +0.0008$, in excellent agreement with the exact value at this coupling, though the comparison is limited since the MAALA correction in Ref.~\cite{Aarts:2026zzr} guarantees accuracy across all $\beta$ values whereas our uncorrected approach degrades at higher couplings.
These results highlight complementary strengths.~The equivariant approach of Ref.~\cite{Aarts:2026zzr} achieves superior accuracy and broader generalisation range, particularly at large $\beta$ and large volumes, owing to the built-in gauge symmetry and the MAALA correction.~Our simpler flat-space approach, while more limited in range, demonstrates that even without explicit gauge equivariance or Metropolis adjustment, a standard diffusion model can learn the correct physics at moderate couplings and lattice sizes close to the training geometry.

However, the two approaches also differ in a way that becomes significant when considering the long-term challenges of lattice field theory.~The MAALA correction requires evaluating the ratio $\exp(-\Delta S_E)$ of real Boltzmann weights at each Metropolis step.~While this guarantees exactness for theories with a real action, the Metropolis correction step itself is not directly applicable to theories with a complex action, such as QCD at finite baryon density, real-time formulations, or theories with a $\theta$-vacuum term, where the Boltzmann weight oscillates.~This limitation applies to any Metropolis-based correction and is not specific to the equivariant framework of Ref.~\cite{Aarts:2026zzr}, which can equally be run without MAALA.~More generally, any uncorrected diffusion approach, whether equivariant or not, never evaluates the action during sampling and relies entirely on the learned score function, making it a potential starting point for theories with a complex action, provided suitable training data can be obtained by alternative methods such as reweighting or complex Langevin dynamics.

Taken together, the equivariant framework of Ref.~\cite{Aarts:2026zzr} and the simpler flat-space approach presented here represent complementary starting points for developing diffusion-based samplers for lattice gauge theories.~The former establishes that high-accuracy generation with broad generalisation is achievable when gauge symmetry and Metropolis correction are built in; the latter shows that even minimal architectures can capture the relevant physics in a regime close to training.~Both approaches contribute to a broader effort aimed at building generative models that can ultimately address problems beyond the reach of traditional methods.

\section{Outlook}
\label{sec:Outlook}

Several directions for extending this work present themselves.~On the architectural side, exploring gauge equivariant networks such as L-CNNs in place of the flat-space U-Net could improve accuracy by encoding symmetries directly rather than requiring the model to learn them from data. For near-term benchmarking in theories with a real action, incorporating a Metropolis correction step analogous to MAALA~\cite{Aarts:2026zzr} would provide a practical way to correct residual model bias; however, developing action-free sampling strategies that maintain accuracy without such corrections remains the more important goal, as these are essential for eventual application to theories with a sign problem.~On the methodological side, training an ensemble of independently initialised models and combining results using random-effects meta-analysis~\cite{DerSimonian1986}, as employed in Ref.~\cite{Aarts:2026zzr}, would provide a principled decomposition of the uncertainty into within-model statistical error and between-model variance, directly probing sensitivity to training stochasticity.

On the physical side, the extension to four-dimensional SU(2) Yang-Mills theory is currently underway.~The key challenges are the increased dimensionality of the link fields (four directions instead of two) and the substantially larger configuration space, which will test both the scalability of the diffusion framework and the effectiveness of physics-conditioned generation in higher dimensions.

For long-term applications involving four-dimensional SU(3) gauge theory, additional challenges arise.~A central one concerns training data efficiency:~our model is trained on $2\times 10^4$ configurations, while Ref.~\cite{Aarts:2026zzr} uses $10^5$ configurations.~Both training sets are relatively large for a two-dimensional theory where HMC is inexpensive, but scaling to four dimensions will require training ensembles on lattice volumes where each configuration could involve $\mathcal{O}(10^6)$ or more link variables.~Producing comparably large training sets may itself become a computational bottleneck, particularly for expensive actions.~Developing architectures and training strategies that can learn effectively from limited training data, for instance by training on smaller lattice extents and generalising to larger volumes, is therefore an important challenge that remains to be investigated.

\begin{acknowledgments}
We thank J.~Bulava, P.~Klein, J.~Gegelia, F.~He, L.~Meng, F.~ Romero-López and W.~Sun for valuable discussions and helpful comments on the manuscript. The calculations of this work were performed on the HPC cluster Elysium of the Ruhr University Bochum, subsidised by the DFG (INST 213/1055-1).
\end{acknowledgments}
\section*{Research Data and Code Access}
The code used for this manuscript is available in Ref.~\cite{sourcefiles_baodong}.

\end{document}